\documentclass[aps,prb,preprint,preprintnumbers,superscriptaddress,amssymb,12pt]{revtex4-1}

\usepackage{graphicx}
\usepackage{dcolumn}
\usepackage{bm}
\usepackage{color}
\usepackage{amsmath,amsfonts,amssymb}
\usepackage{graphicx}
\usepackage{siunitx}
\usepackage[english]{babel}
\usepackage{booktabs,longtable}
\usepackage{natbib}

\newcommand{\abs}[1]{\left|{#1}\right|}

\newcommand{\interpar}[1]{\left({#1}\right)}

\newcommand{\vect}[1]{\mathbf{#1}}
\DeclareMathOperator{\sech}{sech}
\DeclareMathOperator{\csch}{csch}

\begin{document}

\author{William Legrand}
\affiliation{Unit\'e Mixte de Physique, CNRS, Thales, Univ. Paris-Sud, Universit\'e Paris-Saclay, Palaiseau 91767, France}
\author{Jean-Yves Chauleau}
\affiliation{Unit\'e Mixte de Physique, CNRS, Thales, Univ. Paris-Sud, Universit\'e Paris-Saclay, Palaiseau 91767, France}
\affiliation{Synchrotron SOLEIL, L'Orme des Merisiers, 91192, Gif-sur-Yvette, France}
\author{Davide Maccariello}
\author{Nicolas Reyren}
\author{Sophie Collin}
\author{Karim Bouzehouane}
\affiliation{Unit\'e Mixte de Physique, CNRS, Thales, Univ. Paris-Sud, Universit\'e Paris-Saclay, Palaiseau 91767, France}
\author{Nicolas Jaouen}
\affiliation{Synchrotron SOLEIL, L'Orme des Merisiers, 91192, Gif-sur-Yvette, France}
\author{Vincent Cros}
\email{vincent.cros@cnrs-thales.fr}
\author{Albert Fert}
\affiliation{Unit\'e Mixte de Physique, CNRS, Thales, Univ. Paris-Sud, Universit\'e Paris-Saclay, Palaiseau 91767, France}
\title{Hybrid chiral domain walls and skyrmions in magnetic multilayers}

\date{\today}

\begin{abstract}
Noncollinear spin textures in ferromagnetic ultrathin films are currently the subject of renewed interest since the discovery of the interfacial Dzyaloshinskii-Moriya interaction (DMI). This antisymmetric exchange interaction selects a given chirality for the spin textures and allows stabilising configurations with nontrivial topology. Moreover, it has many crucial consequences on the dynamical properties of these topological structures, including chiral domain walls (DWs) and magnetic skyrmions. In the recent years the study of noncollinear spin textures has been extended from single ultrathin layers to magnetic multilayers with broken inversion symmetry. This extension of the structures in the vertical dimension allows very efficient current-induced motion and room-temperature stability for both N\'eel DWs and skyrmions. Here we show how in such multilayered systems the interlayer interactions can actually lead to more complex, hybrid chiral magnetisation arrangements. The described thickness-dependent reorientation of DWs is experimentally confirmed by studying demagnetised multilayers through circular dichroism in x-ray resonant magnetic scattering. We also demonstrate a simple yet reliable method for determining the magnitude of the DMI from static domains measurements even in the presence of these hybrid chiral structures, by taking into account the actual profile of the DWs. The advent of these novel hybrid chiral textures has far-reaching implications on how to stabilise and manipulate DWs as well as skymionic structures in magnetic multilayers.
\end{abstract}

\maketitle

The DMI is a form of antisymmetric exchange\cite{Dzyaloshinskii1958,Moriya1960,Fert1990} that promotes canting between neighbouring magnetic moments. As the DMI in ultrathin films originates from the strong spin-orbit coupling of interfacial atoms neighbouring the magnetic layer\cite{Bogdanov2001,Bode2007}, it is notably found in a ferromagnet (FM) interfaced with two different heavy metals, such as Pt/Co/Ir\cite{Chen2013, Moreau-Luchaire2016}, or in magnetic layers inserted between a heavy metal and an oxide, such as Pt/CoFe/MgO\cite{Emori2013, Emori2014}. This interfacial DMI has a direct influence on the orientation and chirality of the magnetisation textures in these systems, in particular on DWs and magnetic bubbles in perpendicularly magnetised layers. Due to the energy lowering of the N\'eel configuration (internal magnetisation perpendicular to the DW) associated to the interfacial DMI, the demagnetising effects inside the DW that usually favour the Bloch configuration (internal magnetisation along the DW) can be overcome and the N\'eel orientation favoured\cite{Thiaville2012}. Depending on the types of interfacial atoms combined with the FM and on the stacking order, the DMI can also change sign, which determines whether clockwise (CW) or counter-clockwise (CCW) chirality is preferred\cite{Chen2013}. Such stabilisation of chiral magnetic textures also helps to stabilise quasi-punctual solitonic structures called skyrmions, in the present case N\'eel (hedgehog) skyrmions\cite{Nagaosa2013}. In order to obtain stable and compact individual skyrmions at room-temperature, a successful approach has  been to stack up several repeats of an asymmetric combination of ultrathin layers\cite{Woo2016,Moreau-Luchaire2016}. In this way, it is possible to stabilise columnar-shaped skyrmions, in which the increased magnetic volume reinforces their stability against thermal fluctuations up to room-temperature, while still preserving both interfacial perpendicular magnetic anisotropy (PMA) in each ultrathin FM, and required interfacial DMI due to the absence of inversion symmetry of the structure.

Here our purpose is to investigate the thickness-dependent internal profile of chiral magnetic DWs and chiral skyrmions in Pt/Co/Ir and Pt/Co/${\text{AlO}}_{x}$ based multilayers with PMA. The studied multilayers are stackings made of 5 up to 20 repetitions of such trilayers with various magnetic layer thicknesses. Our study combines theoretical predictions, together with a direct experimental observation, of a vertical position dependent reorientation of the chirality inside the magnetisation texture in multilayers. Focusing first on the simplest case of DWs, we predict through micromagnetic simulations not only a significant variation of the DW width through the thickness of the multilayers but, most importantly, a twisting of the internal DW texture along their thickness. In case of a large number of repetitions in the stacking, top and bottom layers then host N\'eel orientations of opposite chiralities (CW and CCW), one in accordance with and the other one opposing the chirality favoured by the interfacial DMI. This reorientation thus results in an uncommon type of chiral composite N\'eel-Bloch DW. Then, we provide an experimental demonstration of the described reorientation of the DW texture in Pt/Co/${\text{AlO}}_{x}$ multilayers by directly observing the DW chirality close to the top surface through circular dichroism in x-ray resonant magnetic scattering. These results involve reconsidering the common implicit assumption of a uniform magnetisation through the thickness of the multilayers. From the better understanding of the actual depth-profile of the DW structure through the thickness, we then describe a novel approach to measure the DMI in such multilayers. Using the experimentally measured size of parallel stripe domains, we show how the DMI strength can be quantitatively extracted for each system by comparing the energies of stripe domains with different sizes in micromagnetic simulations. With the support of analytical calculations, we next propose a simple model which allows to predict the occurrence of DW twisting. We finally discuss how our findings open new opportunities in order to manipulate more efficiently this new type of composite chiral skyrmions in multilayers, through the engineering of the interfacial spin-orbit torques (SOTs).

\section*{Dipolar-field-induced reorientation of magnetic DWs}

In ultrathin magnetic films, the interfacial DMI can be strong enough to stabilise chiral N\'eel DWs. Let us consider a single DW separating two perpendicularly magnetised domains of magnetisation pointing down ($\vect{m}\cdot\vect{z}=-1$) for $x<0$ and up ($\vect{m}\cdot\vect{z}=+1$) for $x>0$. Denoting $\theta(x)=\arccos{\interpar{\vect{m}\cdot\vect{z}}}$ the polar angle and $\psi$ the (fixed) azimuthal angle of the magnetisation inside the DW, the DW energy is given by\cite{Heide2008}
\begin{equation}
\sigma_{\rm{dw}}=A\int_{-\infty}^{+\infty} \interpar{\frac{d\theta}{dx}}^2dx + D\cos\psi\int_{-\infty}^{+\infty} \frac{d\theta}{dx}dx + K\int_{-\infty}^{+\infty} \sin^2\theta{}dx + \sigma_{\rm{dem}}(\psi)
\end{equation}
\noindent where $A$ is the Heisenberg exchange interaction amplitude, $D$ the interfacial Dzyaloshinskii-Moriya interaction, $K$ the uniaxial anisotropy, and $\sigma_{\rm{dem}}$ the demagnetising energy associated to dipolar fields. Without DMI, the $\psi$ dependence of $\sigma_{\rm{dem}}$ favours Bloch type DWs ($\cos\psi=0$). However it is now well known that in the presence of DMI, as comes out from this equation, the energy of the DW can be lower for a N\'eel orientation ($\cos\psi=\pm1$) which causes the reorientation of the internal magnetisation of the DW and explains the stabilisation of N\'eel structures\cite{Bode2007,Heinze2011,Emori2013}. Such N\'eel DWs in ultrathin films exhibit a fixed chirality controlled by the sign of the DMI and, notably, have proven to present extremely interesting current-driven dynamics\cite{Thiaville2012}.

In the recent years, the study of such chiral spin textures has become a very important research field both theoretically and experimentally. Most analyses of the spin configurations in single layers\cite{Thiaville2012} (for DWs and skyrmions) assume a pure N\'eel profile and, for DWs, an energy independent of the DW spacing (uncoupled DWs). However, most often in multilayered stacks the DWs are packed close to each other, meaning that the uncoupled DWs approximation does not apply. Another complexity is that if the demagnetising fields are strong (i.\ e.\ for a large number of repetitions of magnetic layers), a DW structure between Bloch and N\'eel may arise ($0<\psi<\pi/2$). These two effects have been recently considered by Lemesh et al.\ \cite{Lemesh2017}, who have derived a precise analytical model of the demagnetising energies for standard 1D magnetisation profiles. In this model, a classical DW profile with one free internal angle $\psi$ is assumed and the magnetic configuration is parametrised by $\Delta$ (DW width), $\lambda$ (domain periodicity) and $\psi$ only. Knowing the material parameters, a set of three nonlinear equations needs to be solved in order to get the equilibrium value of ($\Delta$,$\lambda$,$\psi$). As we will demonstrate later, this model is very accurate for pure N\'eel walls and allows to estimate the value of $D$, as well as to determine a threshold minimum value of $D$ that prevents internal tilting of the magnetisation inside the DWs towards Bloch configuration.

This ($\Delta$,$\lambda$,$\psi$) model is based on two strong assumptions: (i) the DW has a classical $\arctan$ profile without any $x$ dependence for $\psi$ and (ii) all layers share the same DW profile, that is, without any $z$ dependence. However, it is to be suspected that spin structures not uniform in the $z$ direction can be stabilised in magnetic multilayers, as was suggested by micromagnetic simulations\cite{Montoya2017,Banerjee2017} and imaging of the emergent field of skyrmions, with constraints on the stability in order to reconstruct the magnetic profile\cite{Dovzhenko2016arXiv}. We aim here to demonstrate that in most of the multilayered material systems considered in recent experiments on skyrmions at room-temperature, the actual magnetisation structure through the thickness of the multilayer is indeed more complex. As a result of the competition between the different interactions, a novel type of chiral hybrid magnetic texture is stabilised with some profound consequences on their spin-torque-driven dynamics.

\begin{figure}
\includegraphics[width=17cm, trim= 0cm 4cm 0cm 0cm]{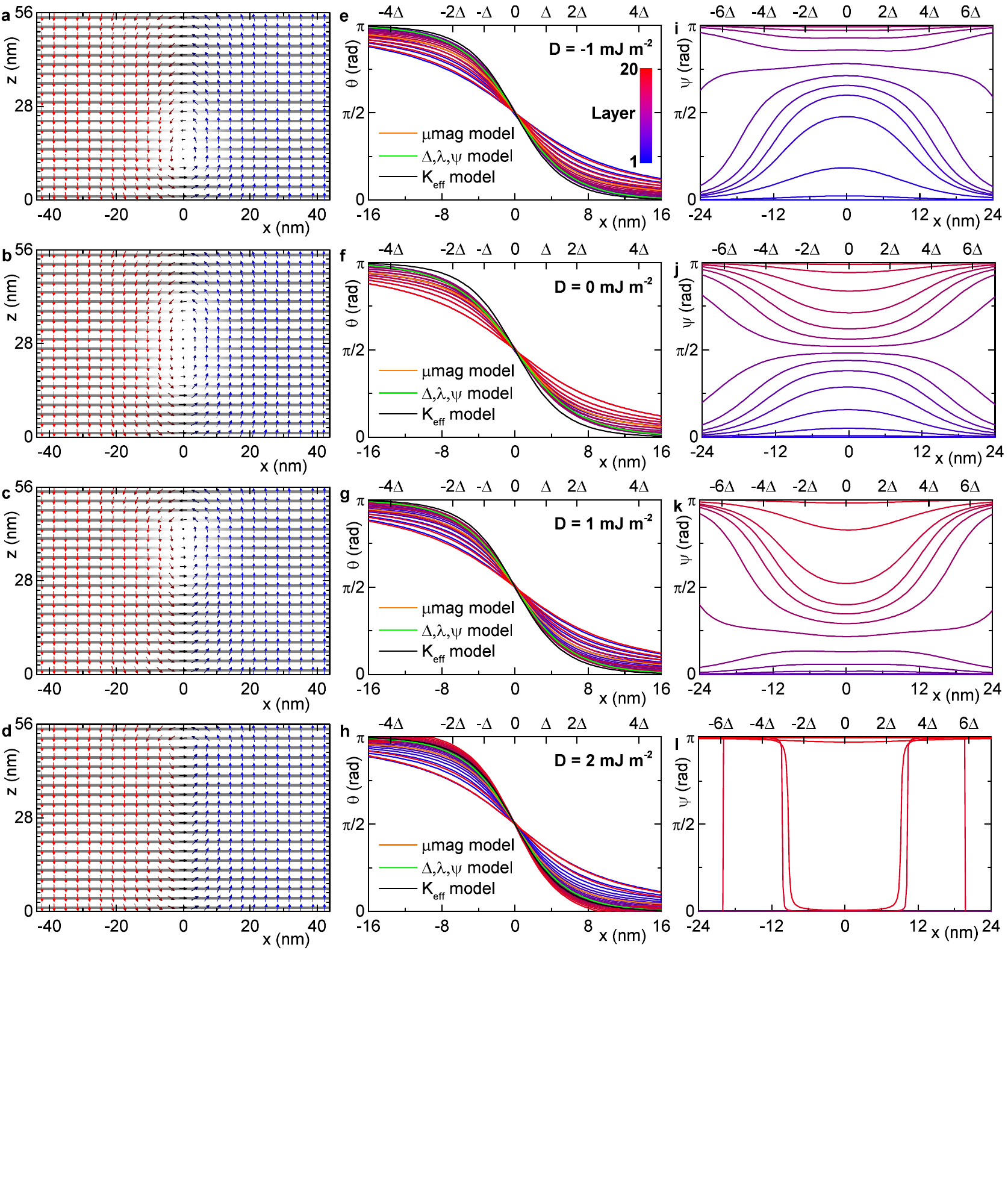}
\caption{a--d.\ Cross-sectional view of a half simulation volume for [X(1)/Co(0.8)/Z(1)$]_{20}$ multilayer with $D=\;$-1.0, 0.0, 1.0 and \SI{2.0}{\milli\joule\per\meter\squared}, respectively. Arrows point in the direction of the magnetisation, $m_z$ is given by the colour of the arrows from red (-1) to blue (+1), while $m_y$ is displayed by the colour of the grid from black (-1) to white (+1). e--h.\ Polar angle $\theta$ inside the DW in each layer for $D=\;$-1.0--\SI{2.0}{\milli\joule\per\meter\squared}. The blue to red lines correspond to layers from bottom to top (see colour scale in e.\;), while the orange line is the average $\theta$ across the thickness. Green and black lines are the profiles as given by the ($\Delta$,$\lambda$,$\psi$) and $K_{\rm{eff}}$ models, respectively. i--l.\ Azimuthal angle $\psi$ inside the DW in each layer for $D=\;$-1.0--\SI{2.0}{\milli\joule\per\meter\squared}. The blue to red lines again correspond to layers from bottom to top.}
\label{fig:DWmodels}
\end{figure}

To support this statement, a first step has been to perform a series of micromagnetic simulations of the DW structure in multilayers. We take the example of a multilayer comprising 20 repetitions of X(\SI{1}{\nano\meter})/Co(\SI{0.8}{\nano\meter})/Z(\SI{1}{\nano\meter}) where X and Z are heavy-metal or oxide non-magnetic spacers, which is typical among multilayers allowing to stabilise skyrmions\cite{Moreau-Luchaire2016,Woo2016,Legrand2017,Litzius2017,Woo2017,Soumyanarayanan2017,Jaiswal2017}, and perform the micromagnetic simulation of the stripe domain configuration in the full geometry including all magnetic layers and non-magnetic spacers (see Methods). In order to study the influence of the DMI on the DW configuration, we vary $D$ to be -1.0, 0.0, 1.0 and \SI{2.0}{\milli\joule\per\meter\squared}, for each of which the DW configuration has been computed. The results of these simulations are compiled in Fig.\ \ref{fig:DWmodels}, in which we present the evolution of the magnetisation profile as a function of the layer position inside the stacking in three different ways. The first column displays cross-sectional views of the magnetisation profile (Figs.\ \ref{fig:DWmodels}a--d), the second column displays the polar angle ($\theta$ angle) of the DW magnetisation across the DW in each layer (Figs.\ \ref{fig:DWmodels}e--h) and the third column displays its azimuthal angle (Figs. \ref{fig:DWmodels}i--l). Each line then corresponds to the different DMI values. 

From these simulations we find that, for a large range of DMI values, (i) the DW internal magnetisation is twisted between N\'eel and Bloch-type along the $z$ direction (i.e. the position of the magnetic layer in the stacking), leading to the formation of what we later call a hybrid chiral DW structure, (ii) the DW width varies through the multilayer thickness and (iii) the $z$ component of the DW magnetisation $m_z$ cannot be simply assumed to follow an $\arctan{}$ function in all layers and none of the usual models allow to fit correctly the DW magnetisation structure. We now proceed with a more detailed description of the results shown in Fig. \ref{fig:DWmodels} with regard to these three points.

First, a clear output from these simulations is that the stabilised DW structures are not corresponding, whatever the value of $D$, neither with pure Bloch walls ($\cos\psi=0$), as usually expected for $D=0$ in single layers (Fig.\ \ref{fig:DWmodels}b), nor with pure N\'eel walls ($\cos\psi=\pm1$) even for $D$ as large as \SI{2.0}{\milli\joule\per\meter\squared} (Fig.\ \ref{fig:DWmodels}d). We further notice that even for intermediate $D=\;\pm$ \SI{1.0}{\milli\joule\per\meter\squared} (Figs.\ \ref{fig:DWmodels}a,c), the DWs do not exhibit an intermediate structure between N\'eel and Bloch with a constant $\psi$ angle of the magnetisation inside the DW, but instead show a continuous variation of this $\psi$ angle between all the magnetic layers of the stacking. Such rather complex internal magnetisation profiles, different from the single layer case, arise in these many-repeats multilayers from the competition between the interfacial DMI and the dipolar interactions between each layers. The resulting configuration can be described by the presence of a mostly Bloch ($\cos\psi$ close to zero) wall part in some of the intermediate magnetic layers, whereas the combined action of the demagnetising fields and DMI fields leads to the stabilisation of mostly N\'eel wall parts in the bottom and top layers. The DW thus has opposite internal magnetisation directions (in some sense, opposite local chiralities) in the top-most and bottom-most layers. This kind of magnetic configuration was known in the absence of DMI for DWs\cite{Labrune1999} and bubbles\cite{Moutafis2006} in magnetic thick layers. Here, we do observe it for a large range of values of the DMI. Interestingly, we notice that the position of the Bloch wall part with respect to the central layer depends on the strength and sign of the DMI. For increasing $\abs{D}$ values, the preference for a single chirality shifts more and more the position of the most Bloch DW (minimal $\abs{\cos\psi}$) up or down depending on the sign of $D$. For example, for $D=\;$\SI{2.0}{\milli\joule\per\meter\squared}, only the two top-most layers contain N\'eel-like DWs with a sense of rotation opposite to all others (for a large negative $D$, the two bottom-most DWs will have an opposite rotation). Only for a large enough value of $D$ ($\abs{D}\geq\;$\SI{2.5}{\milli\joule\per\meter\squared} in the present case), the Bloch-like DWs are excluded and the DW chirality is the same through all layers.

Second, as can be seen from the $\theta(x)$ profiles of Figs.\ \ref{fig:DWmodels}e--h, the DW width varies significantly among the different layers. Again due to dipolar effects, the central, Bloch wall part is more compact than the N\'eel wall parts at top and bottom which extend over a larger width. Overall, the DW widths in the different layers vary significantly, by more than a factor of 2, as can be seen from the slopes at the center of the $\theta(x)$ profiles.

Third, we find that in any individual layer of the stack, the DW profile cannot be properly described by the classical models. On the graphs of Figs.\ \ref{fig:DWmodels}e--h, we add the average over all layers of the $\theta(x)$ profile of the micromagnetic model (orange lines), the DW profile (uniform along $z$) predicted by the so-called $K_{\rm{eff}}$ model (black lines), a model in which the multilayer is treated as an effective magnetic media and the DWs as non-interacting (see Methods for details), and the DW profile (also uniform along $z$) predicted by the ($\Delta$,$\lambda$,$\psi$) model\cite{Lemesh2017} (green lines). It can be clearly seen that the two latter profiles fail to reproduce both the exact shape of the DW, notably in the region of the tail, as well as the DW width. Notably, this difference demonstrates that the actual DW shapes are different from the commonly used $\arctan$ profile. Moreover, as can be seen from the strongly $x$ dependent $\psi(x)$ profiles in Figs.\ \ref{fig:DWmodels}i--l, the Bloch-N\'eel character of the DW in each layer varies across the DW (along $x$), instead of being fixed. As a consequence of the DW being partially Bloch, and only in the intermediate layers (combined $x$ and $z$ dependence of $\psi$), with varying DW widths, the average transverse magnetisation of the DWs is altered and DW energies turn to be significantly different from any simple model predictions. We note that when $D$ exceeds \SI{2.5}{\milli\joule\per\meter\squared} (not shown here), the DWs resemble pure N\'eel DWs, and classical models get accurate, with a good agreement between micromagnetic and classical models.

The main reasons for the discrepancy between the more elaborate micromagnetic simulations and the described analytical models are thus that for the latter no dependence of $\Delta$ along $z$ is considered and that they assume a fixed DW $\arctan$ profile uniform along $z$, thus missing the longer tails in the external layers and the strong variation of $\psi$ along $x$ and $z$. As already described, we instead predict a twisted DW magnetisation through the thickness, constituting a new kind of hybrid chiral DW in DMI systems. The complexity of the DW (or skyrmion, as will be seen later on) profiles in multilayers is well depicted in the curves shown in Figs.\ \ref{fig:DWmodels} i--l. Note that the magnetisation configuration resembles the flux-closure DW arrangements recently observed in magnetic bilayers\cite{Bellec2010, Hrabec2017a}. However, in our multilayer case, the magnetisation evolves more continuously along the $x$ direction, as indicated by the variation of the $\psi$ angle through each layer (see Fig.\ \ref{fig:DWmodels}k for example). 

Therefore, the important difference of hybrid chiral DWs with previous models originates from the competition between the interlayer magnetostatic interactions and DMI, which results in these flux-closed DW configurations that never correspond to DWs with a fixed internal magnetisation angle in between pure Bloch and N\'eel types. We will see later that in our multilayers (but more generally in all the multilayered systems experimentally studied so far), the actual value\cite{Belmeguenai2015} of $D$ is inferior to \SI{2.5}{\milli\joule\per\meter\squared}, which is the threshold value that allows pure N\'eel DWs through all layers in the present case. In consequence, it becomes essential to consider that the actual DW (or skyrmion boundary) textures are similar to the one we have just described. As detailed in the end of this study, this finding has profound implications for the spin-torque-induced manipulation of these composite DWs\cite{Khvalkovskiy2013} or skyrmions\cite{Sampaio2013}.

\section*{Direct observation of the internal twisting of the DWs}

In order to verify experimentally the formation of the above-described composite DWs notably in presence of a significant DMI, we have grown several series of magnetic multilayered stackings by sputtering deposition (see Methods). The structures of the studied multilayers are reported in Table \ref{tab:Samples_list}, starting from the substrate side //, where all thicknesses are given in nanometers and the numbers indexing the brackets are the number of repetitions of the base multilayer. The use of different buffer layers allows us to control  the growth conditions and hence the amplitude of the magnetic anisotropy $H_{k}$ and $M_{s}$ values for similar stacks. The choice of opposite orders of the stackings allows us to get opposite signs of the DMI, while increasing the number of repetitions of the same trilayer base allows us to increase the influence of interlayer dipolar interactions.

\begin{table}
  \scriptsize
  \caption{\label{tab:Samples_list} List of the studied magnetic multilayers investigated. The saturation magnetisation $M_{s}$, the uniaxial anisotropy $H_{k}$, the measured domains periodicity, the estimated DW width, and the comparison of estimations of $D$ with a fixed DW energy ($K_{\rm{eff}}$ model), $(\Delta$,$\lambda$,$\psi)$ model and the present full micromagnetic model are given for each multilayer. Two multilayers are not labelled as they have not been studied by CD-XRMS but only characterised by MFM in order to determine $D$ with the different models.}
  \begin{tabular}{|c|c|c|c|c|c|c|c|c|}
    \hline
     \# & Multilayer stack & $M_{s}$ & $H_{k}$ & $\lambda$ & $\Delta_{\Delta,\lambda,\psi}$ & $D_{\sigma{},K_{\rm{eff}}}$ & $D_{\Delta,\lambda,\psi}$ & $D_{\mathrm{full}}$ \\
      & & (\si{\kilo\ampere\per\meter}) & (\si{\milli\tesla}) & (\si{\nano\meter}) & (\si{\nano\meter}) & (\si{\milli\joule\per\meter\squared}) & (\si{\milli\joule\per\meter\squared}) & (\si{\milli\joule\per\meter\squared}) \\
    \hline
    I & //Pt 10/[\textbf{Ir 1/Co 0.6/Pt 1}$]_{5}$/Pt 3 & 840 & 330 & 167 & 7.27 & -1.46 & -1.43 & -2.30 \\
    II & //Pt 10/[\textbf{Ir 1/Co 0.8/Pt 1}$]_{5}$/Pt 3 & 1229 & 640 & 150 & 4.13 & -2.13 & -2.00 & -2.00 \\
    III & //Pt 11/[\textbf{Co 0.8/Ir 1/Pt 1}$]_{5}$/Pt 3 & 637 & 516 & 278 & 7.08 & 1.48 & 1.45 & 1.37 \\
    IV & //Ta 5/Pt 10/[\textbf{Co 0.8/Ir 1/Pt 1}$]_{5}$/Pt 3 & 683 & 748 & 488 & 5.77 & 1.69 & 1.64 & 1.63 \\
    \hline
    V & //Pt 11/[\textbf{Co 0.8/Ir 1/Pt 1}$]_{10}$/Pt 3 & 637 & 516 & 244 & 6.03 & 1.53 & 1.50 & 1.52 \\
    VI & //Ta 5/Pt 10/[\textbf{Co 0.8/Ir 1/Pt 1}$]_{10}$/Pt 3 & 847 & 1088 & 250 & 4.11 & 2.11 & 2.08 & 2.06 \\
    VII & //Ta 15/Co 0.8/[\textbf{Pt 1/Ir 1/Co 0.8}$]_{10}$/Pt 3 & 957 & 500 & 256 & 5.00 & -1.26 & -1.21 & -1.14 \\
    \hline
    - & //Ta 10/Pt 7/[\textbf{Pt 1/Co 0.6/Al$_{2}$O$_{3}$ 1}$]_{20}$/Pt 3 & 1344 & 469 & 190 & 4.06 & 1.40 & 0.88 & 1.29 \\
    VIII & //Ta 10/Pt 7/[\textbf{Pt 1/Co 0.8/Al$_{2}$O$_{3}$ 1}$]_{20}$/Pt 3 & 1373 & 358 & 175 & 3.72 & 1.20 & 1.20 & 1.01 \\
    - & //Ta 10/[\textbf{Al$_{2}$O$_{3}$ 1/Co 0.6/Pt 1/}$]_{20}$/Pt 7 & 1120 & 332 & 131 & 4.79 & -1.91 & -1.93 & -1.94 \\
    IX & //Ta 10/[\textbf{Al$_{2}$O$_{3}$ 1/Co 0.8/Pt 1/}$]_{20}$/Pt 7 & 1245 & 228 & 131 & 4.33 & -1.76 & -1.78 & -1.69 \\
    \hline
  \end{tabular}
\end{table}

To get evidence of the existence of the predicted DW twisting we have measured the circular dichroism in X-ray resonant magnetic scattering (CD-XRMS), allowing us to directly probe the magnetisation sense of rotation of the top layers of the stackings\cite{Chauleau2017arXiv}. This experimental approach has proven to be a relevant and straightforward approach to probe the chiral aspect of closure magnetic domains in FePd\cite{Durr1999} as well as the topological winding and chirality in $\text{Cu}_2\text{OSeO}_3$\cite{Zhang2017a,Zhang2017b}. In particular, we have recently demonstrated that the DW chirality (CW or CCW) could be unambiguously determined by assessing the dichroic pattern sign in opposite stackings of Pt/Co/Ir multilayers\cite{Chauleau2017arXiv}. For the present experiments, the energy is set at the $L_3$ Co edge and the angle of incidence of the X-rays has been chosen to be $18.5^{\circ}$, corresponding to the first diffraction peak of the multilayer. Under these conditions, X-rays are mainly sensitive to the first \SI{15}{\nano\meter}, which corresponds to only 4 to 5 repetitions from the top of the multilayered structure that predominate in the measured XRMS signal (see Methods). Before performing the XRMS experiments, all the samples have been demagnetised in order to reach a magnetic ground state composed either of labyrinthine alternating up and down magnetised domains or parallel stripe domains depending on the chosen demagnetisation procedure (see Methods).

\begin{figure}
\includegraphics[width=17cm, trim= 0cm 10cm 0cm 0cm]{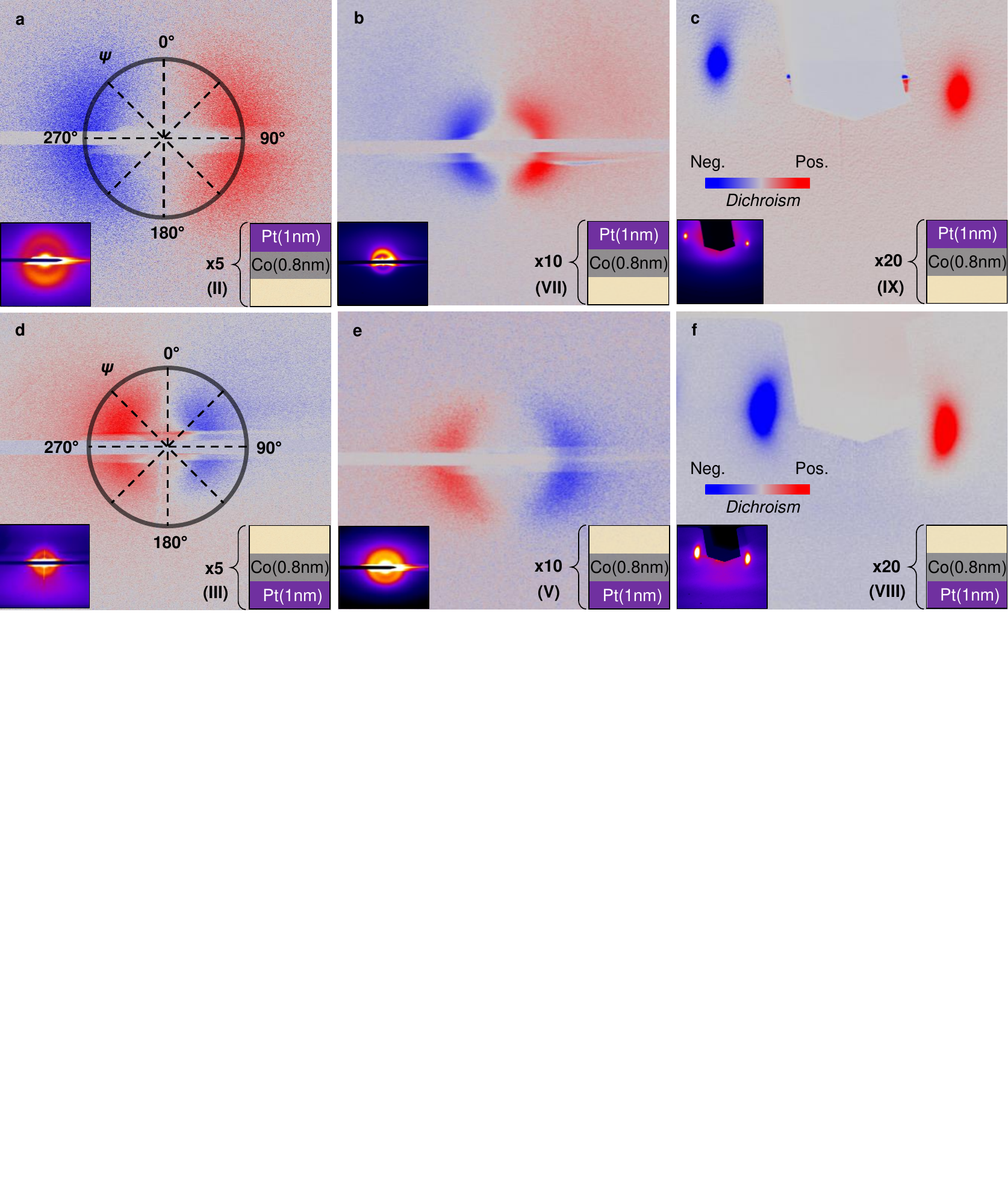}
\caption{Circular dichroism analysis of different multilayer stacking configurations with $D<0$ and a.\ 5 repetitions (sample II), b.\ 10 repetitions (sample VII) and c.\ 20 repetitions (sample IX); with $D>0$ and d.\ 5 repetitions (sample III), e.\ 10 repetitions (sample V) and f.\ 20 repetitions (sample VIII). The dichroism is normalised for each image and indicated by the colour scale from blue (negative) to red (positive). Left insets are the corresponding sum (CL+CR) images evidencing the magnetic distribution ordering. Right insets present schemes of the studied stackings.}
\label{fig:XRMSdata}
\end{figure}

In Figs.\ \ref{fig:XRMSdata}a--c, we display the XRMS dichroism patterns (diffraction patterns with unpolarised X-rays are shown as insets) recorded for the multilayers labelled (II), (VII) and (IX) (with 5, 10 and 20 repetitions, respectively). These three multilayers are the ones having the Pt layer on top of the FM ($D<0$, see Table \ref{tab:Samples_list}). In the second line (see Figs.\ \ref{fig:XRMSdata}d--f), the dichroism patterns are the ones obtained for the multilayers labelled (III), (V) and (VIII) (also with 5, 10 and 20 repetitions, respectively) that have the Pt layer below the FM ($D>0$, see Table \ref{tab:Samples_list}). We emphasise that for the first two columns, the multilayers have been demagnetised with an oscillating out-of-plane field, leading to a labyrinthine domain configuration that generates rings in the diffraction patterns. On the contrary, for multilayers (IX) and (VIII) shown in Figs.\ \ref{fig:XRMSdata}c,f, the demagnetisation procedure was done applying an in-plane field, leading to parallel stripe domains configuration and resulting in diffraction spots appearing on both sides of the specular peak. The corresponding MFM images are shown in the Supplementary Material.

Important conclusions can be drawn from this series of XRMS patterns. First, because the dichroic signal is in all cases maximum along the $90^{\circ}$-$270^{\circ}$ axis, it can be concluded that all the DWs in these different multilayers share the same type of DW in surface, corresponding to N\'eel DWs as we demonstrated recently\cite{Chauleau2017arXiv}, instead of tilted DWs in between N\'eel and Bloch configurations. Second, we clearly observe that for all samples with Pt on top of the FM (Figs.\ \ref{fig:XRMSdata}a--c), a positive (red colour in our convention) dichroism at $90^{\circ}$ is found. This means that the top few layers exhibit a CW N\'eel DW chirality whatever the number of repetitions in the multilayers, which is the DW configuration that can be expected for this sign of the DMI. On the opposite, for the first two multilayers with Pt below the FM (Figs.\ \ref{fig:XRMSdata}d,e) the dichroic signal is reversed (maximum positive at $270^{\circ}$). It indicates an opposite, CCW N\'eel DW chirality for 5 and 10 repetitions, again as expected from the sign of the DMI. A striking observation is then that the dichroism pattern recorded for the 20-repeats multilayer with Pt below the FM (shown in Fig.\ \ref{fig:XRMSdata}f) is again positive at $90^{\circ}$ and thus indicates a CW N\'eel chirality in surface, which is opposite to what is expected for a positive-DMI-driven chirality. As we have explained previously, this apparent discrepancy is a direct fingerprint of the competition between interlayer dipolar fields and DMI: due to the 20 repeats in the multilayer, the impact of the dipolar fields is large enough to impose a reversal of the chirality of the DWs in the top layers. Note that we are able to observe this dipolar-field induced chirality reversal only for sample (VIII) and not for multilayer (IX), because for $D<0$ ($D>0$), the DMI field is opposite to the in-plane component of the dipolar field in the bottom (top) layers (see the magnetic configurations in Figs.\ \ref{fig:DWmodels}a-d). As a consequence, for sample (IX) the local DW chirality reversal occurs in the bottom layers, which are not the ones probed by the X-rays in our XRMS configuration, and thus the reorientation is not observed. The chirality reversal does not occur for the 5 and 10-repeats multilayers because the strength of the dipolar fields relative to the DMI fields in these multilayers is not strong enough, as will be shown later. These results thus directly demonstrate the dipolar-field-induced twisting of DWs in magnetic multilayers as described in the previous section. 

\section*{Estimation of $D$ with hybrid chiral DW textures}

Being able to get a reliable quantitative estimation of the interfacial DMI amplitude has been the subject of numerous studies in the last couple of years as this parameter is crucial to understand and control both the statics and dynamics of chiral DWs and/or skyrmions. Different approaches have been proposed. In single magnetic layers, the interfacial DMI has been experimentally accessed using Brillouin light spectroscopy \cite{Di2015, Belmeguenai2015}, spin-wave spectroscopy \cite{Lee2016}, chirality-induced asymmetric DW propagation \cite{Emori2014,Hrabec2014} or asymmetric magnetisation reversal \cite{Pizzini2014,Han2016,Kim2017a}. However, these different methods appear to be not reliable in case of interfacial DMI in multilayered systems. A first reason is that dipolar fields and magnetic couplings between layers shall influence significantly the spin-wave propagation and thus the analysis of BLS spectra. Moreover, because dense labyrinthic magnetic domains are formed at low fields, it thus prevents the observation of the asymmetric reversal needed to estimate the DMI. This is why in multilayered systems, analysis of domain spacing in the demagnetised state or as a function of the perpendicular magnetic field has been proposed to estimate the magnitude of DMI \cite{Woo2016,Moreau-Luchaire2016}. This approach is based on the fact that the DW periodicity $\lambda$ is the result of the balance between domains demagnetisation energy and DW energy\cite{Kooy1960}, the latter being dependent on $D$, as described above. Assuming that the observed demagnetised state is very close to the state of minimum energy given parallel stripe domains, we can then determine the DW energy $\sigma_{\rm{dw}}$ and deduce a rough estimation of the DMI magnitude $\abs{D}$ with the so-called $K_{\rm{eff}}$ model (see Methods). This characterisation method has led to the evidence of a significant DMI in magnetic multilayers with broken inversion symmetry \cite{Woo2016, Moreau-Luchaire2016}. 

It has however been pointed out recently that such measurements can be largely erroneous when neglecting stray-field effects on the DW size and spacing, so that a more comprehensive model is required for multilayers\cite{Lemesh2017}. When dipolar interactions become significant but the DW internal configuration remains uniform, a more detailed model such as the ($\Delta$,$\lambda$,$\psi$) is more accurate\cite{Lemesh2017}. Nevertheless, we now suspect that complex DW or skyrmion structures can arise in magnetic multilayers depending of the relative strengths of dipolar interactions and DMI, which calls for a more careful analysis of their spin textures\cite{Banerjee2017, Montoya2017, Dovzhenko2016arXiv}.  

We show here how the existence of hybrid chiral DWs induced by dipolar fields has a large impact on the evaluation of $D$ in multilayers. To get an accurate estimation even in the presence of the hybrid chiral DW structures we have identified, our procedure is then to rely on the complete micromagnetic simulations as exemplified above. We vary the domains period in the simulation by changing the number of cells around the measured $\lambda$ of the parallel stripe domains and then relax the system for different values of $D$. The extracted value $D_{\rm{full}}$ is $D$ for which the energy density of the simulated system is minimum at $\lambda$ (see Methods). The extracted $D$ values for the different multilayers are listed in Table \ref{tab:Samples_list}, in which we also compare $D_{\rm{full}}$ with the values of $D$ estimated with the $K_{\rm{eff}}$ and ($\Delta$,$\lambda$,$\psi$) models. Although all models give consistent values for pure N\'eel DWs, we note significant differences as soon as at least one layer has a reversed chirality. 

We emphasise that the measurement of the domain periodicity $\lambda$ is error-free from the size of the MFM imaging probe\cite{Bacani2016arXiv}, and that the sensitivity to local defects is reduced due to the averaging effect of imaging over large sample areas. Thus, we believe that the most straightforward mean to quantify the DMI in multilayers remains so far to find it from the measurement of $\lambda$ in the ground state. Finally, another interesting aspect of this approach using the periodicity of stripe domains is that aligning parallel stripe domains with varying directions allows to measure the DMI along different directions, which can give information on a potential anisotropic DMI in materials which have a crystalline structure allowing different DMI vectors along their different crystalline axes. We did not find anisotropic DMI in our samples.

\section*{Criterion for dipolar-field induced twisting of DW chirality}

\begin{figure}
\includegraphics[width=8.5cm, trim= 0cm 0cm 0cm 0cm]{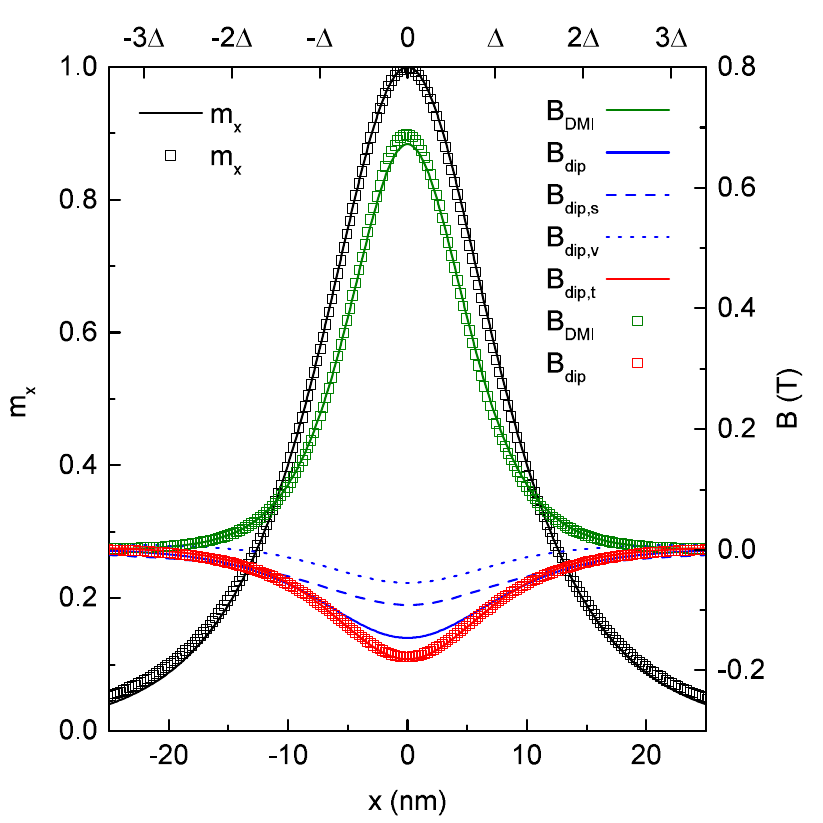}
\caption{DW profile (black, left scale) and analytical model for estimation of DMI (green, right scale) and dipolar (red, right scale) fields for the top layer of (III) [Pt(1)/Co(0.8)/Ir(1)$]_{5}$. The squares are the result of micromagnetic minimisation of the energy, while the lines are the fields obtained from the model. The dashed and dotted blue lines are respectively surface and volume charge contributions to the interaction dipolar field (solid blue line). The red line is the total dipolar field from the model (obtained adding the intralayer demagnetising field to the interlayer interaction dipolar field). The parameter $\Delta$ has been adjusted for the magnetisation arctan analytical profile to fit the micromagnetic profile.}
\label{fig:Dipfields}
\end{figure}

In the following, our objective is to establish a simple criterion describing the occurrence of twisted chiral DWs in magnetic multilayered systems. For that we follow an approach opposite to what we have just described as we now study the DW structure considering a known value of $D$. Considering parallel domains and by making the assumption that all DWs are N\'eel (with $D>0$), we can study the stability of this pure N\'eel configuration. To this aim we find under these assumptions the horizontal components of the dipolar- and DMI-induced fields inside the top layer (see Methods). We then compare the magnitude of the average action of both these fields on the DW profile, $\mathcal{A}_{\rm{dip}}$ and $\mathcal{A}_{\rm{dmi}}$, over a total width of $6\Delta$ around the center of the DW (see Methods). As an example, we show these deduced field profiles (green and blue lines for DMI fields and interlayer dipolar interaction fields, respectively, right scale) for our multilayer (III) [Pt(1)/Co(0.8)/Ir(1)$]_{5}$ in Fig.\ \ref{fig:Dipfields}, together with the DW horizontal magnetisation component $m_x(x)$ profile (black curve, left scale). The dipolar interaction field $B_{\rm{dip}}$ (solid, blue line) is obtained by summing surface (dashed, blue line) and volume (dotted, blue line) magnetic charges contributions. The small difference between the total field $B_{\rm{dip,t}}$ and the dipolar interaction field with the other layers $B_{\rm{dip}}=B_{\rm{dip,S}}+B_{\rm{dip,V}}$ comes from the horizontal component of the self-demagnetising field of the top layer itself, which has no influence on the reorientation and that we do not include in the comparison of field actions. These analytically calculated field and magnetisation profiles match the ones obtained from micromagnetic simulations (shown as squares of the corresponding colours in Fig.\ \ref{fig:Dipfields}) under the assumptions mentioned just above.

\begin{figure}
\includegraphics[width=8.33cm, trim= 0cm 0cm 0cm 0cm]{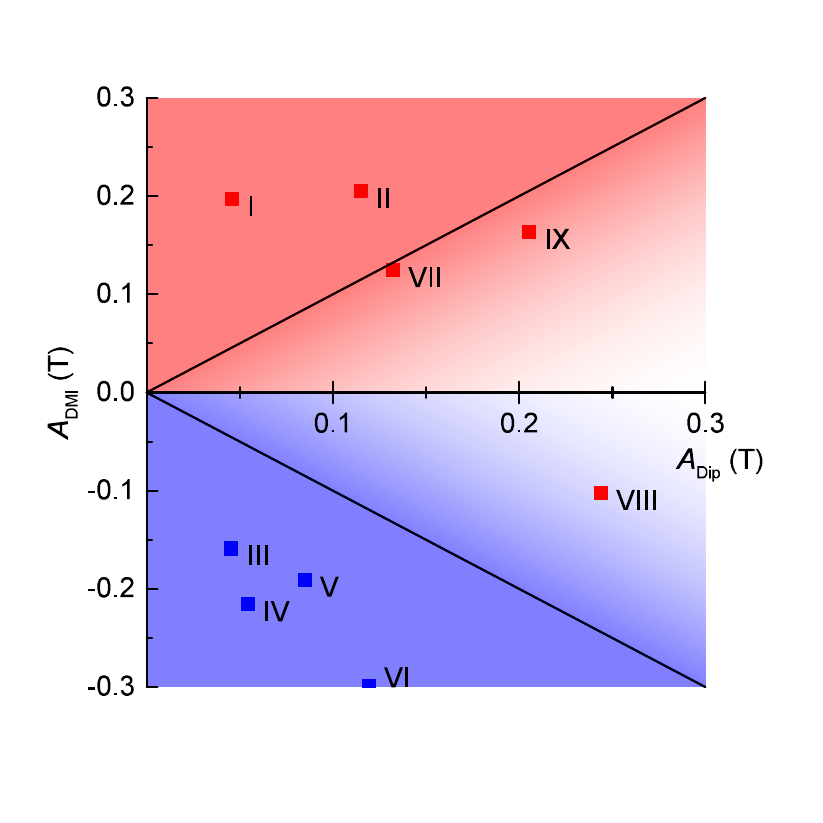}
\caption{Diagram comparing $\mathcal{A}_{\rm{dip}}$ and $\mathcal{A}_{\rm{dmi}}$ for each multilayer that has been characterised by CD-XRMS. When $\abs{\mathcal{A}_{\rm{dip}}}<\abs{\mathcal{A}_{\rm{dmi}}}$ pure N\'eel DWs are stabilised, and red (blue) indicates CW (CCW) chirality. The gradient areas correspond to $\abs{\mathcal{A}_{\rm{dip}}}>\abs{\mathcal{A}_{\rm{dmi}}}$ with more and more pronounced reorientation into flux-closure DWs. Coloured squares indicate the chirality as it has been observed by CD-XRMS for each sample labelled by roman numbers, where red stands for CW chirality and blue stands for CCW chirality.}
\label{fig:XRMSsamples}
\end{figure}

When $\abs{\mathcal{A}_{\rm{dip}}}>\abs{\mathcal{A}_{\rm{dmi}}}$, for $D>0$ ($D<0$) the top layer (bottom layer) in-plane component of the magnetisation inside the DW will reverse due to dipolar fields, so that a pure N\'eel DW throughout the whole stack is definitively not stable, whereas for $\abs{\mathcal{A}_{\rm{dip}}}<\abs{\mathcal{A}_{\rm{dmi}}}$ N\'eel DWs of the same chirality can be stabilised in all layers. We can thus predict whether the DWs will reorientate or not, as shown in Fig.\ \ref{fig:XRMSsamples}. The areas filled with a uniform colour correspond to $\abs{\mathcal{A}_{\rm{dip}}}<\abs{\mathcal{A}_{\rm{dmi}}}$ in which case pure N\'eel-like DWs are stabilised. The colour indicates the preferred DW chirality imposed by the DMI. When this condition is not fulfilled, i.\ e.\ , $\abs{\mathcal{A}_{\rm{dip}}}>\abs{\mathcal{A}_{\rm{dmi}}}$, corresponding to the gradient areas, the DW reorientation into flux-closure DWs leading to the stabilisation of hybrid chiral DWs shall occur. The twisting is more and more pronounced when $\abs{\mathcal{A}_{\rm{dip}}}$ becomes stronger relative to $\abs{\mathcal{A}_{\rm{dmi}}}$, which is signified by the progressively lighter background colour. In order to compare these predictions with our experiments, we also include in Fig.\ \ref{fig:XRMSsamples} the experimental observation of the DW chirality of the top layers determined by CD-XRMS in all our different multilayers. The colour of the squares labelled (I-IX) indicate the top-surface DW sense of rotation, blue corresponding to CCW and red to CW chiralities. As described above, the DW surface chirality for multilayer (VIII) is found in agreement with the prediction, i.\ e.\ , the corresponding square is red in a blue gradient area, meaning that the observed chirality is CW, due to dipolar fields, whereas the DMI favours a CCW chirality. We further notice that for multilayers (VII) and (IX), the DW reorientation that shall occur in the bottom layers is indeed not observed by our enhanced surface sensitive technique. This can be seen in Fig.\ \ref{fig:XRMSsamples}, as the corresponding squares are red in a red gradient area. We believe that these predictions are important as they allow to easily obtain for a given set of magnetic parameters $M_{s}$, $K_{u}$, $A$ and multilayer geometry, an approximation of the threshold DMI value $D_u$ that ensures a unique DW or skyrmion chirality inside the whole stacking. We will see in the last section that knowing whether twisting of the magnetisation chirality occurs or not is crucial as the DW or skyrmion dynamics are strongly modified depending on their actual spin textures. 

\section*{Consequence on dynamics of skyrmions in multilayers}

\begin{figure}
\includegraphics[width=17cm, trim= 0cm 0.5cm 0cm 0cm]{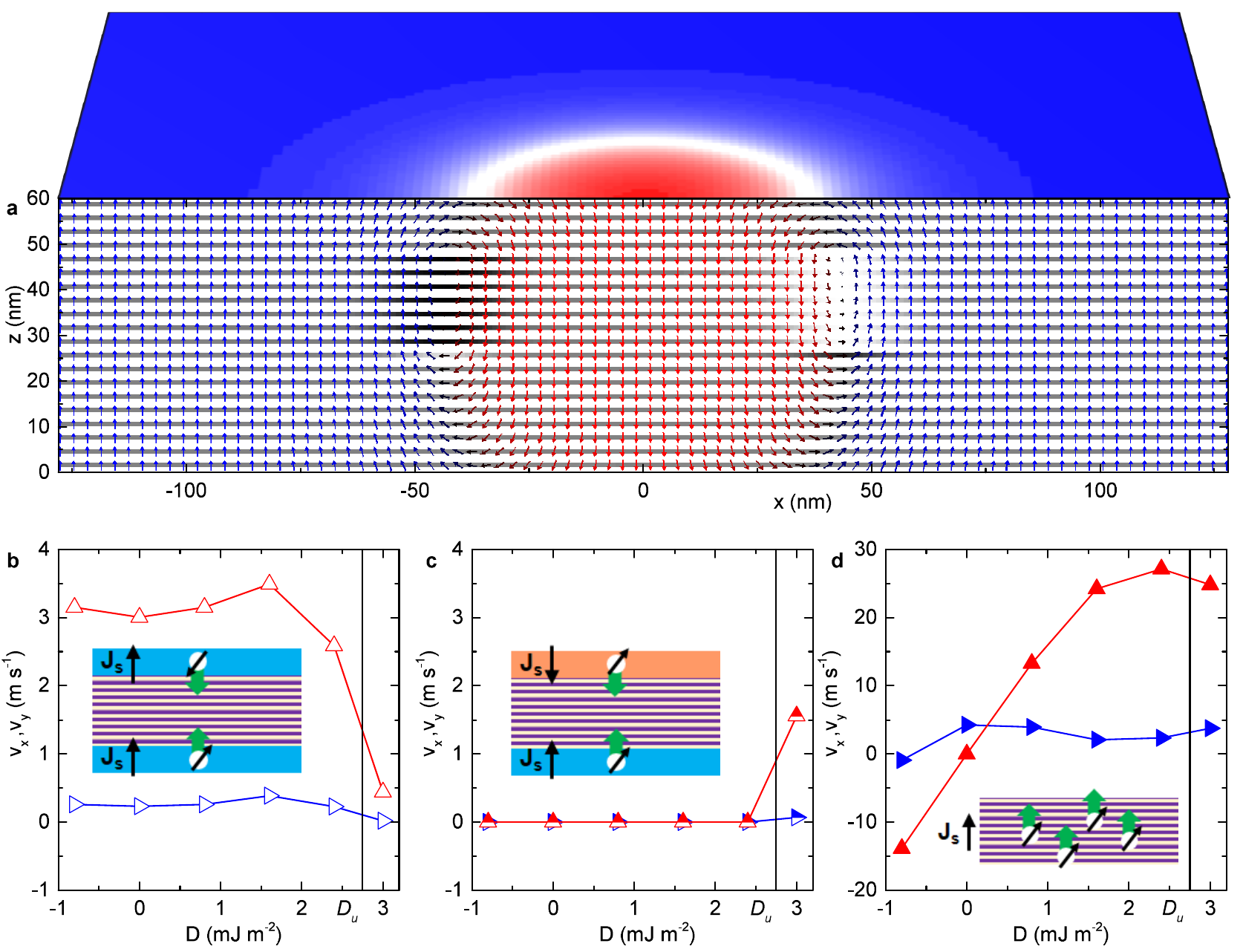}
\caption{a.\ Cut view of the simulation volume for a [X(1)/Co(1)/Z(1)$]_{20}$ multilayer with $D=\;$\SI{0.8}{\milli\joule\per\meter\squared}. Arrows point in the direction of the magnetisation, $m_z$ is given by the colour of the arrows from red (-1) to blue (1), while $m_y$ is displayed by the colour of the grid from black (-1) to white (+1). The $m_z$ component in the top layer is represented in perspective view by the colour from red (-1) to blue (1) b-d.\ Skyrmion velocities for different values of $D$ and geometries. Right/up pointing blue/red triangles stand for horizontal/transverse velocity components, obtained for b.\ opposite injection in bottom-most/top-most layers, c.\ identical injection in bottom-most/top-most layers and d.\ uniform injection. The injection geometry is depicted by the inset in each case.}
\label{fig:SKmodels}
\end{figure}

Stackings of ultrathin magnetic layers have been the subject of a large research effort in the last years for demonstration of room-temperature stabilisation of small magnetic skyrmions. Moreover, it has already been shown\cite{Woo2016,Hrabec2017a,Woo2017} that the interfacial spin-orbit torques can be used in order to move efficiently these skyrmions. As aforementioned, the existence of a complex spin texture through the different magnetic layers of a stacking due to the competition between interlayer dipolar fields and DMI fields is not only expected for DWs (and observed as we have demonstrated here) but should be equivalent for skyrmions in multilayers, as it has been also pointed out in other very recent works\cite{Montoya2017,Dovzhenko2016arXiv}. The fact that columnar skyrmions stabilised in multilayers, which is the favourite strategy to make them stable at room temperature, possess a twisting of their chiral spin texture through their thickness shall strongly alter their dynamics. However, to our knowledge this crucial issue has never been properly considered so far. We show in Fig.\ \ref{fig:SKmodels}a an example of the actual skyrmion profile in a multilayer of structure [X(1)/Co(1)/Z(1)$]_{20}$ with $D=\;$\SI{0.8}{\milli\joule\per\meter\squared}, obtained by micromagnetic modeling similar as before (see Methods). Equivalently to what happens for DWs, the skyrmion profile exhibits a progressive reorientation from CCW N\'eel to Bloch through its thickness, and finally to CW N\'eel chirality in the topmost magnetic layers. Moreover the skyrmion diameter also evolves depending on the layer position in the stack, being larger in the central layers and smaller in the external layers. Such skyrmions with N\'eel caps were recently predicted and thoroughly modeled by micromagnetic simulations for amorphous, \SI{50}{\nano\meter} Gd/Fe thick layers\cite{Montoya2017}. With our findings, we demonstrate that the stabilisation of such hybrid chiral skyrmions occurs in multilayered stackings of ultrathin magnetic and non-magnetic layers with strong interfacial DMI. In our multilayers, the different magnetic layers are indeed exchange decoupled due to Pt/Ir and Pt/Al$_{2}$O$_{3}$ spacers, which enhance reorientation effects as compared to exchange-coupled bulk materials like Gd/Fe\cite{Montoya2017}. In the following, we show that through a precise engineering of the interfacial spin torques in stacked multilayers, new possibilities can be anticipated for achieving a good control of the current-driven skyrmion dynamics.

To this aim, we have simulated the current-induced dynamics of isolated hybrid chiral skyrmions in an extended multilayer similar to the one shown in Fig.\ \ref{fig:SKmodels}a, for different values of the DMI and different spin injection geometries. Note that we only consider here damping-like torques originating from vertical spin currents, which can be provided for example by the spin Hall effect and a current flowing along the $x$ direction. The resulting skyrmion velocities are reported  in Fig.\ \ref{fig:SKmodels}b-d. In Fig.\ \ref{fig:SKmodels}b, we first present the case in which the torque is applied only in the first bottom and the first top layers and the injected spin polarisation is opposite in these two layers, for example, with the multilayer enclosed between two layers of the same heavy metal. We find that both the longitudinal and transverse velocities remain constant for moderate values of $D$ (with transverse velocity larger than longitudinal velocity, as it is already known, due to the gyrotropic motion of the topological skyrmions\cite{Legrand2017}) but then drops to zero when $\abs{D}>D_u$, corresponding to the critical minimal value of $D$ above which skyrmions with a unique chirality across all layers are stabilised. The velocity drops because skyrmions of opposite chiralities are driven in opposite directions for identical polarisations of injected spins, but in the same direction for opposite spin injections. As a consequence the driving forces on the bottom CCW and top CW N\'eel skyrmion layers add up in the range of $D$ for which the chirality twist is present and cancel out for a uniform skyrmion chirality. In Fig.\ \ref{fig:SKmodels}c, we present the results for the opposite case where the injected spins have identical polarisations in the bottom-most and top-most layers, for example, with the multilayer enclosed between heavy metals of opposite spin Hall angles. In that case we find that the skyrmion motion is completely canceled up to $D_u$ because the hybrid complex spin texture leads to opposite skyrmion chiralities in bottom and top layers. The motion occurs only for $\abs{D}>D_u$, in which case top and bottom layers chiralities and injected spins are identical. Finally, for a uniform current injection (Fig.\ \ref{fig:SKmodels}d, for example, with a Pt layer adjacent to the bottom of each ferromagnetic layer), we find that the transverse ($v_y$) velocity is roughly proportional to $D$ up to $D_u$ (note the reduced velocity in cases of only external layers injection as compared to uniform injection within all layers, as fewer spins are injected in total). Again it is because skyrmions of opposite chiralities are driven in opposite directions, which implies that the global motion is related to the balance of CCW and CW N\'eel skyrmions layers. Even if $D$ does not directly affect the velocity of a given structure, as $D$ controls the amount of layers with CCW and CW N\'eel orientations, it also controls the overall velocity. We also notice that in this case, the driving forces compensate at $D=0$, leading to a zero global $v_y$ . The small $v_x$ component arises from the Bloch part of the hybrid chiral skyrmion. Interestingly, this case corresponds to the experimental studies of skyrmion dynamics reported in Refs.\ \onlinecite{Woo2016, Litzius2017} and provide a simple explanation why a very large (and hard to achieve) interfacial DMI is mandatory to achieve fast skyrmion motion in stacked multilayered systems in comparison to the case of single layers or multilayers with few repeats. The high value of $D_u\approx\;$\SI{2.8}{\milli\joule\per\meter\squared} for 20 repetitions of X/Co(\SI{1}{\nano\meter})/Z is above all reported values in multilayers. This shows that pure N\'eel skyrmions in such structures are very unlikely to occur. However, $D_u$ can be lowered to realistic values in 20 repetitions of a thinner ferromagnetic layer such as in Pt/Co(\SI{0.6}{\nano\meter})/${\text{AlO}}_{x}$ or in structures for which $M_{s}$ is lower. Finally, this series of simulations thus provides guidelines on how to design multilayers with many repetitions hosting skyrmions, and how to engineer the interfacial spin torques in order to achieve a fast motion when hybrid chiral skyrmions are present in these multilayers.

\section*{Outlook}

The presented skyrmion dynamics simulations open new perspectives for the engineering of multilayered magnetic materials. Using such techniques, it will be possible to better control DW and skyrmion motion with spin-orbit torques in asymmetric multilayers and even in most common PMA symmetric stacks, such as Pt/Co and Pd/Co, which exhibit only low DMI or even no significant DMI. Notably, injecting spin Hall effect spin currents from the same material below and above the multilayer will lead to motion, as the chirality of the structure reverses from bottom to top. Moreover, as we have shown with the system displayed in Fig.\ \ref{fig:DWmodels}, the balance between the number of layers and the value of $D$ in the structure actually controls the position of the Bloch DW within the multilayer. It is thus possible to control the overall chirality of the structures by engineering the base multilayer repetition number, thus tuning the ratio of CCW N\'eel, Bloch, and CW N\'eel DWs. This would allow to control the direction of the motion in the case of a uniform spin current injection as well as the chirality-related properties of DW and skyrmion motion. Moreover, our simulations of current-induced motion show the dynamical stability of the twisted chiral skyrmions as they are robust even under current injection, which can thus be used for spin-current-induced motion.

With the help of our description, it is possible to understand and predict the occurrence of twisting of DWs and skyrmions chirality in multilayers. Notably, the common picture of a constant internal tilt inside the DW or constant chirality inside the skyrmion and through the different layers is revealed not to be valid. We have experimentally demonstrated this reorientation effect with a surface-sensitive X-ray diffraction technique, allowing to observe the surface chirality of the DW configurations. Such complex structures strongly modify the way the DMI should be quantified in magnetic multilayers.

These results highlight the importance of advanced engineering of spin-orbit related interfacial properties, combining PMA, SOT and DMI in multilayered systems to promote the stabilisation and the fast dynamics of ultrasmall skyrmions at room-temperature, which are needed for any type of potential applications.

\section*{Methods}

\subsection*{Samples fabrication and characterisation}
All series of multilayers were grown by dc and rf magnetron sputtering at room-temperature on oxidised silicon substrates, after deposition of a Ta or Pt buffer as described in Table \ref{tab:Samples_list} and capped with Pt($\geq3$ \si{\nano\meter}) in order to prevent oxidation. The saturation magnetisation $M_{S}$ and effective perpendicular anisotropy field $H_{\rm{eff}}=H_{\rm{K}}-M_{S}/2$ of each multilayer are obtained by SQUID and alternating gradient magnetometry measurements.

\subsection*{Demagnetisation procedeure and MFM imaging}
We have first compared the domain periodicity in the labyrinthine and aligned parallel domains configurations (obtained after out-of-plane and in-plane demagnetisation, respectively), measured by magnetic force microscopy (MFM, see Methods) in lift mode for our multilayer (VIII) [Pt(1)/Co(0.8)/${\text{AlO}}_{x}$(1)$]_{20}$.  For both in-plane and out-of plane demagnetisation procedures, the maximum oscillating field was set higher than the anisotropy field in the multilayers, i.e. $B\geq\;$\SI{1.2}{\tesla}. Then the field was set to oscillate between positive and negative values with a decay factor of 0.5\% until \SI{0.5}{\milli\tesla} is reached. For the in-plane demagnetisation procedure the multilayer had a small tilt angle ($\sim5^\circ$) with the field to favour multiple reversals of the magnetisation everywhere in order to get closer to a parallel stripes ground state. The extraction of the mean domain periodicity $\lambda$ after Fourier transform reveals that it can differ significantly, by up to 20\%, between the two demagnetisation procedures. All rigorous models of magnetic domains in multilayers have however been derived for parallel stripe domains\cite{Kooy1960, Draaisma1987, Kaplan1993}. Even if the difference in the domain width for the two demagnetised state configurations may be very small\cite{Lemesh2017} in the ideal case, this deviation precludes the use of perpendicularly demagnetised domains for the estimation of the DMI, as only the values derived from the parallel stripe domains periodicity provide consistent results. The multilayers were imaged with Asylum Low Moment tips in double pass, tapping mode followed by lift mode at \SI{20}{\nano\meter} height, at room-temperature.

\subsection*{XRMS measurements}
XRMS experiments have been carried out at the SEXTANTS beamline of the SOLEIL synchrotron. The diffraction was measured in reflectivity conditions for circularly left (CL) and right (CR) polarisations of the incident X-ray beam. The photon energy was set at Co $L_3$ edge (778.2 eV) using the RESOXS diffractometer. The diffracted X-rays were imaged on a square CCD detector covering $6.1^{\circ}$ at the working distance of this study. All the images have been geometrically corrected along the $Q_x$-direction in order to account for the projection effect related to the photon incidence angle of $18.5^{\circ}$. The sum of the images obtained with CL and CR polarised light gives rise to a diffraction pattern around the specular beam (which was blocked by a beamstop to avoid saturation of the CCD) in the reciprocal plane ($Q_x$,$Q_y$). Denoting $I_{CL}$ and $I_{CR}$ the intensities collected by the CCD for CL and CR polarised incident lights, the circular dichroism (CD) of the scattering signal is defined as $(I_{CL}-I_{CR})/(I_{CL}+I_{CR})$. 

\subsection*{Domain wall energies - $K_{\rm{eff}}$ model}
For uncoupled, independent DWs, separating domains of size $\lambda/2$ much larger than the DW width $\Delta=\sqrt{A/K}$, there exist two straightforward approximations for $K$ and thus for the DW energy\cite{Mougin2007}. For single and ultrathin layers (of thickness $t\ll\Delta$), the demagnetising fields favour in-plane magnetisation inside the DW so that the anisotropy affecting the DW is the effective perpendicular anisotropy $K=K_{u}-\mu_{0}M_{s}^2/2$. On the contrary, for thick layers ($t\gg\Delta$), the in-plane alignment inside the DW is disfavoured and $K=K_{u}+\mu_{0}M_{s}^2/2$. However, these approximations cannot be applied in the intermediate thickness regime, which is the relevant case for magnetic multilayers such as the ones considered here, where both $\Delta$ and $t$ are in the order of \SI{10}{\nano\meter}. Following Ref. \onlinecite{Mougin2007}, we can refine the formula of the DW anisotropy $K$ by assuming that for calculating demagnetising fields roughly, the DW can be considered as a monodomain magnetic body of width $2\Delta$, height $t$ and infinite length. Due to the $\arctan$ profile of the DW, an elliptic shape is a good approximation. Inserting the demagnetising factors of the elliptic cylinder\cite{Osborn1945}
\begin{eqnarray*}
N_x&=&\frac{t}{t+2\Delta}\\
N_z&=&\frac{2\Delta}{t+2\Delta}
\end{eqnarray*}
\noindent in the effective anisotropy $K$ gives a simple expression which allows to find the DW width by solving
\begin{equation*}
\Delta=\sqrt{\frac{A}{K_{u}+\frac{\mu_{0}M_{s}^2}{2}\frac{(t-2\Delta)}{(t+2\Delta)}}}
\end{equation*}
By summing the contributions of exchange, DMI, anisotropy and demagnetising fields in the elliptic body\cite{Thiaville2012}, the DW energy is then
\begin{equation*}
\sigma_{\rm{dw}}=2A/\Delta+2K_{u}\Delta-\pi\abs{D}+\frac{\mu_{0}M_{s}^2}{2}2\Delta \left(\frac{t-2\Delta}{t+2\Delta}\right)
\end{equation*}
\noindent Concerning the demagnetising energy arising from the domains, in the zero width (as $\lambda \gg \Delta\sim 0$) model we get\cite{Kooy1960}
\begin{equation*}
\epsilon_{\rm{demag}}=2\mu_{0}M_{s}^{2}\frac{\lambda}{t}\frac{1}{\pi^{3}}\sum_{n\geqslant1 ,\textrm{odd}}^{\infty} \frac{1}{n^3}\left (1-\mathrm{e}^{-2\pi{}nt/\lambda}\right ) 
\end{equation*}
\noindent which allows one to find $\sigma_{\rm{dw}}$ from the observed $\lambda$. Indeed, by minimising the total energy $\epsilon_{\rm{tot}}=2\sigma_{\rm{dw}}/\lambda+\epsilon_{\rm{demag}}$ relative to $\lambda$ we get
\begin{equation*}
\sigma_{\rm{dw}}=\mu_{0}M_{s}^{2}\frac{\lambda^{2}}{t}\frac{1}{\pi^{3}}\sum_{n\geqslant1 ,\textrm{ odd}}^{\infty} \frac{1}{n^3}\left (1-\mathrm{e}^{-2\pi{}nt/\lambda}-\frac{2\pi{}nt}{\lambda}\mathrm{e}^{-2\pi{}nt/\lambda } \right ) 
\end{equation*}
\noindent which allows one to find $D$ by equaling it to the previous expression of $\sigma_{\rm{dw}}$ as we know $M_{s}$, $K_{u}$, and $A$ estimated to be \SI{10}{\pico\joule\per\meter}. $A$ was obtained by determining the Curie temperature from temperature-dependent SQUID measurements. To apply this $K_{\rm{eff}}$ model, the multilayer is treated as an effective magnetic medium filled with diluted moments. Note that this assumption is valid as long as the periodicity of the stack is not significantly larger than the DW size\cite{Lemesh2017}, which is always verified for the samples considered here.

\subsection*{Domain wall energies - micromagnetic simulations}
In order to find the strength of the DMI in the stripe domains configuration without making assumptions on the DW profiles, we performed micromagnetic simulations with the Mumax3 solver\cite{Vansteenkiste2014} in a 3D mesh accounting for the full geometry of the multilayers. The simulation volume is $\lambda\pm{}d\lambda\;\times$ \SI{32}{\nano\meter} $\times \;Np$, respectively, in the $x$, $y$ and $z$ directions. Two DWs separating up, down and up domains are initialised at $(-\lambda\pm{}d\lambda)/4$ and $(\lambda\pm{}d\lambda)/4$, which corresponds to the $\lambda\pm{}d\lambda$ periodicity of the stripes. Periodic boundary conditions inclusive of the periodic stray fields calculated for $64\times64$ identical neighbors in the $x$ and $y$ directions were introduced. The $x$ cell size was \SI{0.25}{\nano\meter} for the XRMS multilayers series. The $z$ cell size was \SI{0.2}{\nano\meter}, or \SI{0.1}{\nano\meter} when required by the values of layer spacing $p$ and thickness $t$. Simulations are performed at \SI{0}{\kelvin}.

Given the experimental value of $\lambda$, the simulation is initialised with its DWs having a $\psi=45^{\circ}$ in-plane tilt of internal moments for $x$ sizes $\lambda -$\SI{2}{\nano\meter},$\lambda -$\SI{1}{\nano\meter},$\lambda$,$\lambda +$\SI{1}{\nano\meter},$\lambda +$\SI{2}{\nano\meter}. Each system is relaxed in order to find the ground state energy density $\epsilon(\lambda)$, so that we get the local value of $d\epsilon/d\lambda$ at $\lambda$. Performing this operation for various values of $D$ allows to find $D_{\rm{full}}$ such that $d\epsilon/d\lambda=0$ by interpolation.

\subsection*{Domain wall fields - Analytical derivation}
To find the dipolar field in the top layer of the multilayer, we have to find the solution for the potential $\phi$ of Laplace equation $\nabla^2\phi=\rho_V$ with $\rho_V$ the volume charges and boundary conditions related to surface charges $\rho_S$. As was done in Lemesh et al.\cite{Lemesh2017} we separate volume and surface magnetic charges contributions. We consider $\lambda$-periodic stripe domains in the $x$ direction (uniform along $y$) and approximate the DW profile by the arctan profile. We note $t$ the magnetic layer thickness, $p$ the multilayer periodicity and $N$ the total number of layers. Assuming a perfectly N\'eel DW in all layers, we can obtain the DW width $\Delta$ from the $(\Delta$,$\lambda$,$\psi)$ model.

We first solve for a single layer $\nabla^2\phi=0$ with $\partial\phi/\partial{}z(x,\pm{}t/2^{-})=\partial\phi/\partial{}z(x,\pm{}t/2^{+})\pm\rho_S(x)$.  As $\rho_S(x)$ corresponds to the charges of two opposite, alternate DW profiles (up to down and down to up) localised every $\lambda/2$, we can write\cite{Lemesh2017}
\begin{equation*}
\rho_S(x)=\sum_{k=-\infty}^\infty f_S(x)\ast\delta(x-k\lambda) + \sum_{k-\infty}^\infty -f_S(x)\ast\delta(x-k\lambda-\lambda/2)
\end{equation*}
where $f$ corresponds to a single DW profile $f_S(x)=M_{s}m_{z}(x)=M_{s}\tanh{\interpar{x/\Delta}}$, that is, by combining all Dirac functions and swapping the derivatives in the convolution product,
\begin{equation*}
\rho_S(x)=\frac{f_S'(x)}{2}\ast{}g(x)
\end{equation*}
with
\begin{equation*}
g(x)=\begin{cases} 
	1 & x\in\left[k\lambda ; \lambda/2+k\lambda\right[ \\
	-1 & x\in\left[\lambda/2+k\lambda;\lambda+k\lambda\right[.
	\end{cases}
\end{equation*}
As the magnetic charge distribution is $\lambda$ periodic we can decompose it in Fourier series 
\begin{equation*}
\rho_S(x)=\sum_{k=-\infty}^\infty\bar{\rho_S}(k){\rm e}^{-\frac{2\pi{}kx}{\lambda}}
\end{equation*}
and solve
\begin{equation*}
\nabla^2\bar{\phi}(k,z)=\frac{\partial^2\bar{\phi}(k,z)}{\partial{}z^2}-\interpar{\frac{2\pi{}k}{\lambda}}^2\bar{\phi}(k,z)=0
\end{equation*}
with $\partial\bar{\phi}/\partial{}z(k,\pm{}t/2^{-})=\partial\bar{\phi}/\partial{}z(\pm{}t/2^{+})\pm\bar{\rho_S}$. By properties of the Fourier transform defined as
\begin{equation*}
\bar{f}(\xi)=\frac{1}{\sqrt{2\pi}}\int{}f(x){\rm{e}}^{-i\xi{}x}dx
\end{equation*}
we know that
\begin{equation*}
\bar{\rho_S}=\overline{f_S'/2\ast{}g}=\sqrt{2\pi}\interpar{\overline{f_S'}\bar{g}}/2=\sqrt{2\pi}i\xi\overline{f_S}\bar{g}/2
\end{equation*}
so that for positive integer k,
\begin{align*}
\bar{\rho_S}(\frac{2\pi{}k}{\lambda})&=\frac{\sqrt{2\pi}i\pi{}k}{\lambda}\left[-iM_{s}\Delta\sqrt{\frac{\pi}{2}}\csch{\interpar{\frac{\pi^2\Delta{}k}{\lambda}}}\right]\left[-\frac{4i}{\sqrt{2\pi}k}\sin^2\interpar{\frac{k\pi}{2}}\right] \\
&=-i\sqrt{\frac{\pi}{2}}\frac{4\pi{}M_{s}\Delta}{\lambda}\sin^2\interpar{\frac{k\pi}{2}}\csch{\interpar{\frac{\pi^2\Delta{}k}{\lambda}}}.
\end{align*}

Using the boundary conditions to solve Laplace equation\cite{Draaisma1987} above the layer we find for $z>t/2$
\begin{equation*}
\phi(x,z)=\sum\limits_{k=1}^\infty \frac{2M_{s}\Delta}{k}\sin^2{\interpar{\frac{\pi{}k}{2}}}\csch{\interpar{\frac{\pi^2\Delta{}k}{\lambda}}}\sinh{\interpar{\frac{\pi{}kt}{\lambda}}}\sin{\interpar{\frac{2\pi{}kx}{\lambda}}}{\rm{e}}^{-\frac{2\pi{}kz}{\lambda}}
\end{equation*}
with $x=0$ in the center of the DW between down and up domains. For a multilayer there are $N$ layers located at $p/2+kp$ with $0\leq{}k\leq{}N-1$. In the top layer, the interlayer interaction field will then be the sum of all other layers stray fields
\begin{equation*}
B_{\rm{dip,S}}(x)=-\mu_0\frac{\partial}{\partial{}x}\left[\sum\limits_{n=1}^{N-1}\phi(x,np)\right]
\end{equation*}
that is by grouping the exponent sum 
\begin{align*}
B_{\rm{dip,S}}(x)&=-\mu_0\frac{\partial}{\partial{}x}\left[\sum\limits_{k=1,\rm{odd}}^\infty \frac{2M_{s}\Delta}{k}\csch{\interpar{\frac{\pi^2\Delta{}k}{\lambda}}}\sinh{\interpar{\frac{\pi{}kt}{\lambda}}}\sin{\interpar{\frac{2\pi{}kx}{\lambda}}}\frac{{\rm{e}}^{-\frac{2\pi{}kp}{\lambda}}-{\rm{e}}^{-\frac{2\pi{}kNp}{\lambda}}}{{1-\rm{e}}^{-\frac{2\pi{}kp}{\lambda}}}\right] \\
&=-\mu_0\sum\limits_{k=1,\rm{odd}}^\infty \frac{4\pi{}M_{s}\Delta}{\lambda}\csch{\interpar{\frac{\pi^2\Delta{}k}{\lambda}}}\sinh{\interpar{\frac{\pi{}kt}{\lambda}}}\cos{\interpar{\frac{2\pi{}kx}{\lambda}}}\frac{{\rm{e}}^{-\frac{2\pi{}kp}{\lambda}}-{\rm{e}}^{-\frac{2\pi{}kNp}{\lambda}}}{{1-\rm{e}}^{-\frac{2\pi{}kp}{\lambda}}}
\end{align*}
which can be determined numerically.

We can now solve for a single layer $\nabla^2\phi=\rho_V(x)$ for $\abs{z}<t/2$ and $\nabla^2\phi=0$ for $\abs{z}>t/2$, with continuity of $\partial\phi/\partial{}z$ at $z=\pm{}t/2$. We have again
\begin{equation*}
\rho_V(x)=\sum_{k=-\infty}^\infty f_V(x)\ast\delta(x-k\lambda) + \sum_{k-\infty}^\infty -f_V(x)\ast\delta(x-k\lambda-\lambda/2)
\end{equation*}
with
\begin{equation*}
f_{V}(x)=-M_{s}\nabla\cdot\vect{m}=-M_{s}\frac{\partial{}m_x}{\partial{}x}=\frac{M_{s}}{\Delta}\frac{\tanh{\interpar{x/\Delta}}}{\cosh{\interpar{x/\Delta}}}
\end{equation*}
the volume charges for a single DW. Similar to $\rho_S$ we get $\rho_V(x)=[f_V'(x)\ast{}g(x)]/2$ and for positive integer k,
\begin{align*}
\bar{\rho_V}(\frac{2\pi{}k}{\lambda})&=\overline{f_V'/2\ast{}g}(k)=\sqrt{2\pi}i\interpar{2\pi{}k/\lambda}\overline{f_V}\bar{g}/2 \\
&=\frac{\sqrt{2\pi}i\pi{}k}{\lambda}\left[iM_{s}\Delta\frac{2\pi{}k}{\lambda}\sqrt{\frac{\pi}{2}}\sech{\interpar{\frac{\pi^2\Delta{}k}{\lambda}}}\right]\left[-\frac{4i}{\sqrt{2\pi}k}\sin^2\interpar{\frac{k\pi}{2}}\right] \\
&=-i\sqrt{\frac{\pi}{2}}\frac{-8\pi^2M_{s}\Delta{}k}{\lambda^2}\sech{\interpar{\frac{\pi^2\Delta{}k}{\lambda}}}\sin^2\interpar{\frac{k\pi}{2}}.
\end{align*}
Using the boundary continuity to solve Laplace equation above the layer we find for $z>t/2$
\begin{equation*}
\phi(x,z)=\sum\limits_{k=1}^\infty \frac{2M_{s}\Delta}{k}\sin^2{\interpar{\frac{\pi{}k}{2}}}\sech{\interpar{\frac{\pi^2\Delta{}k}{\lambda}}}\sinh{\interpar{\frac{\pi{}kt}{\lambda}}}\sin{\interpar{\frac{2\pi{}kx}{\lambda}}}{\rm{e}}^{-\frac{2\pi{}kz}{\lambda}}
\end{equation*}
so that
\begin{equation*}
B_{\rm{dip,V}}(x)=-\mu_0\sum\limits_{k=1,\rm{odd}}^\infty \frac{4\pi{}M_{s}\Delta}{\lambda}\sech{\interpar{\frac{\pi^2\Delta{}k}{\lambda}}}\sinh{\interpar{\frac{\pi{}kt}{\lambda}}}\cos{\interpar{\frac{2\pi{}kx}{\lambda}}}\frac{{\rm{e}}^{-\frac{2\pi{}kp}{\lambda}}-{\rm{e}}^{-\frac{2\pi{}kNp}{\lambda}}}{{1-\rm{e}}^{-\frac{2\pi{}kp}{\lambda}}}
\end{equation*}

One still has to consider the self-demagnetising field of the top layer itself. The surface charges distribution does not contribute to the $z$-average of the field as it generates a field antisymmetric in z. However the volume charges contribution must be considered. Solving again the Laplace equation but inside the magnetic layer ($\abs{z}<t/2$) we find
\begin{equation*}
B_{\rm{dip}}(x,z)=\mu_0\sum\limits_{k=1,\rm{odd}}^\infty \frac{4\pi{}M_{s}\Delta}{\lambda}\sech{\interpar{\frac{\pi^2\Delta{}k}{\lambda}}}\cos{\interpar{\frac{2\pi{}kx}{\lambda}}}\left[\cosh{\interpar{\frac{2\pi{}kz}{\lambda}}}{\rm{e}}^{\frac{\pi{}kt}{\lambda}}-1\right],
\end{equation*}
that we can average between $z=-t/2$ and $z=t/2$ to obtain the horizontal component of the self-interaction field of the top layer
\begin{equation*}
B_{\rm{dip,self}}(x)=\mu_0\sum\limits_{k=1,\rm{odd}}^\infty \frac{4\pi{}M_{s}\Delta}{\lambda}\sech{\interpar{\frac{\pi^2\Delta{}k}{\lambda}}}\cos{\interpar{\frac{2\pi{}kx}{\lambda}}}\left[\frac{1-{\rm{e}}^{-\frac{2\pi{}kt}{\lambda}}}{2}\frac{\lambda}{\pi{}kt}-1\right].
\end{equation*}
The dipolar interlayer interaction pushing to reverse the DW is then
\begin{equation*}
B_{\rm{dip}}(x)=B_{\rm{dip,S}}(x)+B_{\rm{dip,V}}(x)
\end{equation*}
while the total dipolar field is finally
\begin{equation*}
B_{\rm{dip,t}}(x)=B_{\rm{dip,S}}(x)+B_{\rm{dip,V}}(x)+B_{\rm{dip,self}}(x).
\end{equation*}

The N\'eel internal field along $x$ of the DW can be described by\cite{Thiaville2012}
\begin{equation*}
B_{\rm{dmi}}(x)=\frac{2D}{M_{s}}\frac{\partial{}m_{z}}{\partial{}x}=\frac{2D}{M_{s}\Delta}\sech^2{\interpar{\frac{x}{\Delta}}}
\end{equation*}
as we approximate it to a classical tan profile.

In order to compare the strengths of $B_{\rm{dip}}$ and $B_{\rm{dmi}}$, we evaluate their actions locally by
\begin{equation*}
\mathcal{A}=\frac{1}{6\Delta}\int_{-3\Delta}^{3\Delta}B(x)m_x(x)dx
\end{equation*}
that corresponds to the in-plane rotation driving force.

\subsection*{Skyrmion velocities - micromagnetic simulations}
In order to find the potential current-induced velocities for different skyrmion chirality conditions and spin current geometries, we performed micromagnetic simulations in the full geometry for the multilayer [X(1)/Co(1)/Z(1)$]_{20}$. Here, Co(1) was chosen in order to reduce the simulation grid size. The simulation volume is \SI{256}{\nano\meter} $\times$ \SI{256}{\nano\meter} $\times \;Np$, respectively, in the $x$, $y$ and $z$ directions. Periodic boundary conditions inclusive of the periodic stray fields calculated for $3\times3$ identical neighbors in the $x$ and $y$ directions were introduced. All cell sizes were \SI{1}{\nano\meter}. Before current is applied the configuration is relaxed from an initial Bloch skyrmion with CCW wall internal magnetisation. The current density is then modeled as a fully polarised (along $y$) vertical spin current of current density $J=$ \SI{2.5e10}{\ampere\per\meter\squared}. The Gilbert damping $\alpha$ was set to 0.1 and the out-of-plane field to $B=200$, or \SI{300}{\milli\tesla} (for $D>D_{u}$). Due to this required increase of external field to confine the skyrmion to a stable circular shape for high $D>D_{u}$, the skyrmion size is reduced and thus velocity as well\cite{Legrand2017}, without affecting the discussed quantitative behavior. Simulations were performed at \SI{0}{\kelvin}. 

\section*{Author contributions}
W.L., J.-Y.C., N.R., V.C. and N.J. conceived the project; W.L. deposited the multilayers with the help of S.C.; W.L performed the magnetic characterisation, and together with D.M. and K.B., performed the MFM measurements of domains spacing; J.-Y.C., N.J., N.R. and V.C. performed the XRMS experiments; W.L. developed the analytical derivation part; W.L. and N.R. performed the micromagnetic simulations; all authors contributed to the analysis and interpretation of the experimental results and to the writing of the manuscript.

\begin{acknowledgments}
We gratefully acknowledge I. Lemesh and F. B{\"{u}}ttner for sharing their manuscript\cite{Lemesh2017} while it was still in press. Financial support from FLAG-ERA SoGraph (ANR-15-GRFL-0005) and European Union grant MAGicSky No.\ FET-Open-665095 is acknowledged.
\end{acknowledgments}

\bibliography{DMI_stripes}

\end{document}